\def\softd{{\leavevmode\setbox1=\hbox{d}%
\hbox to 1.05\wd1{d\kern-0.4ex{\char039}\hss}}}%cstocs
\def\softt{{\leavevmode\setbox1=\hbox{t}%
\hbox to \wd1{t\kern-0.6ex{\char039}\hss}}}%cstocs
\def\softl{l\kern-0.45ex\raise0.1ex\hbox{'}\kern-0.10ex}%cstocs
\def\softL{L\kern-0.8ex\raise0.1ex\hbox{'}\kern0.1ex}%cstocs
\documentstyle[a4,12pt,epsfig,psfig,amssymbols]{article}
\begin{document}
\thispagestyle{empty}
\leftline{{\it Nuclear Centre of Charles University, Prague}\hfill
PRA-HEP-95/5}
\leftline{\it Institute of Physics, Prague}
\begin{center}\mbox{\epsfig{file=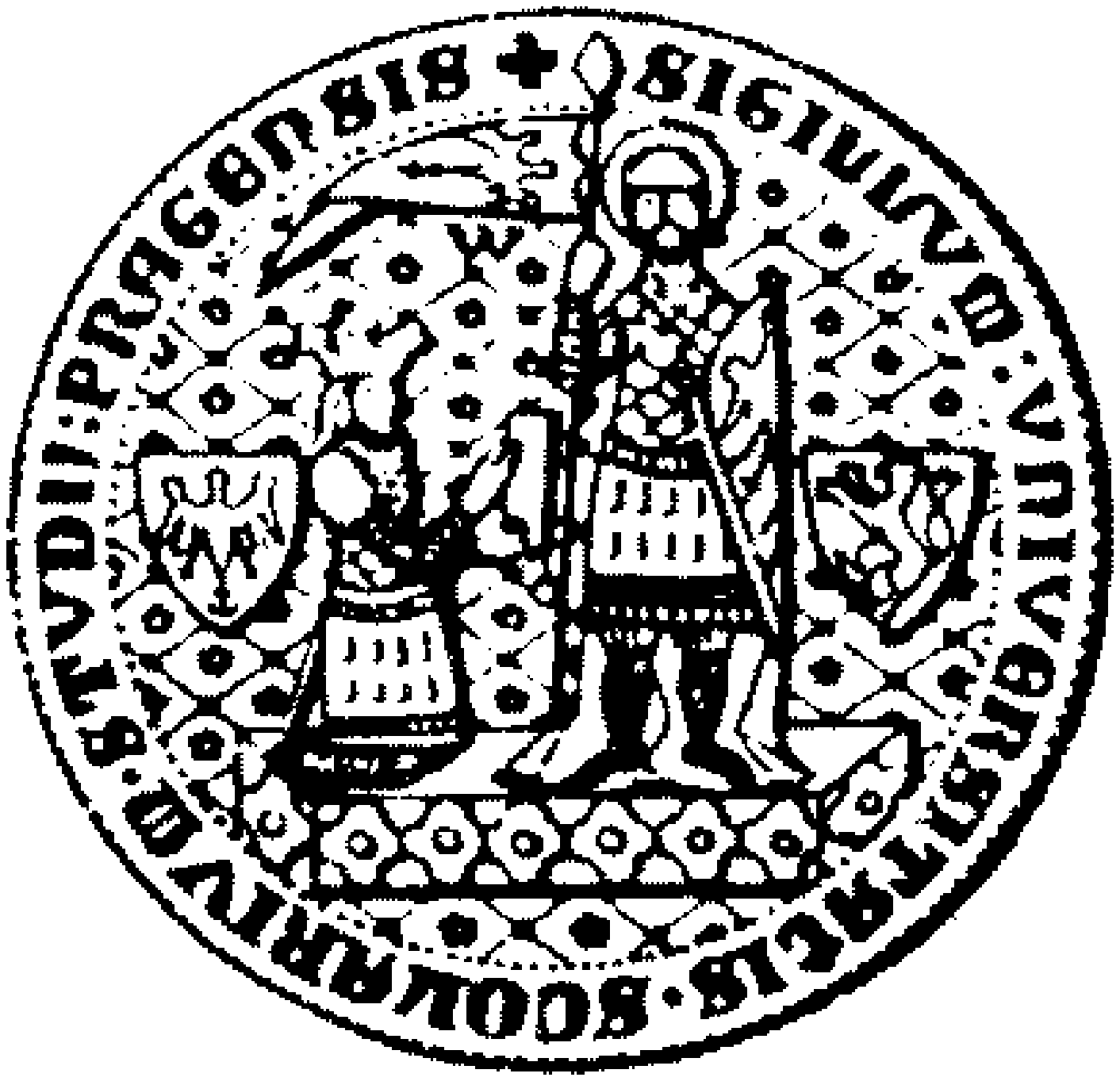,height=5cm}}\end{center}

\begin{center}
{\LARGE Measurement of spin-dependent total cross-section difference
$\Delta\sigma_T$  in  neutron-proton scattering at 16
MeV}

\vspace{1cm}

{\bf J. Bro\v{z}, J. \v{C}ern\'y, Z.
Dole\v{z}al\footnote{electronic mail address
dolezal@hp01.troja.mff.cuni.cz, fax +42-2-8576-2434},
G.M. Gurevich\footnote{on leave of absence from
Institute for Nuclear Research, Russian Academy of Sciences,
60th October anniversary prospect 7A,  MOSCOW, 117312 Russia},
M. Jir\'asek, \\P. Kub\'{\i}k,
A.A. Lukhanin\footnote{on leave of absence from
Kharkov Institute of Physics and Technology, Academy Str. 1,
KHARKOV, 310108 Ukraine},
J. \v{S}vejda and I. Wilhelm}

\vspace{0.1cm}

Nuclear Centre, Charles University, V~Hole\v{s}ovi\v{c}k\'ach 2,
PRAGUE 8,
\\ CZ-180 00 Czech Republic

\vspace{0.4cm}

{\bf N.S. Borisov, Yu.M. Kazarinov,
B.A. Khachaturov, E.S. Kuzmin, V.N. Matafonov, A.B.
Neganov, I.L. Pisarev, Yu.A. Plis and \\ Yu.A. Usov}

\vspace{0.1cm}

Joint Institute for Nuclear Research, Laboratory of Nuclear Problems,
DUBNA, Moscow Region, 141980  Russia

\vspace{0.4cm}

{\bf  M. Rotter and B. Sedl\'ak}

\vspace{0.1cm}

Department of Low Temperature Physics,  Faculty of Mathematics and
Physics, Charles University, V~Hole\v{s}ovi\v{c}k\'ach 2, PRAGUE 8,
CZ-180 00 Czech Republic

\vspace{0.4cm}

11 July 1995

\end{center}

\newpage
\thispagestyle{empty}
\begin{abstract}
        A~new measurement of $\Delta\sigma_T$   for polarized neutrons
transmitted through a polarized proton target at 16.2 MeV has been made.
A polarized neutron beam was obtained from the
$^{3}\rm{H}(d,\vec n)^{4}\rm{He}$ reaction;
proton polarization over 90\% was achieved in a frozen spin target
of 20 cm$^3$ volume.
The measurement yielded the value
$\Delta\sigma_T=(-126\pm21\pm14)$ mb. The result of a
simple phase shift analysis for the $^3S_1-^3D_1$
 mixing parameter
$\epsilon_1$ is presented and compared  with the theoretical
potential model predictions.

\end{abstract}

\vspace{1cm}
\vspace{4cm}
\begin{center}
Submitted to Zeitschrift f\"ur Physik A
\end{center}
\newpage
%\tableofcontents

\setcounter{page}{1}
\section{Introduction} \label {intro}

For the description of nucleon-nucleon ($NN$) forces
semiphenomenological  potential models are
used~\cite{{lac},{mac},{sto}}.
However there are some discrepancies between various
parametrizations,
particularly in the isosinglet $n-p$ system. For the critical
evaluation of potential models measurements of many
variables in a wide range of energies are necessary. From the
existing reviews (e.g. \cite{lehar}) of $NN$ scattering
experiments one can see that there are numerous measurements of
$pp$ scattering (contributing to the description of $I=1$
system), but only sparse data exists for the $n-p$ system.
This situation is even worse at low energies (below 100 MeV),
especially for polarized experiments.  However the importance of
spin-dependent observables for the nuclear interaction theory
development is evident.

  Optical theorem shows that cross-sections (unlike
other observables) depend linearly on the scattering amplitudes
so they give us a more direct information for the understanding
of the $NN$ forces.
The lack of data
in the low energy region is visible from the Fig. 1, where the
situation in the measurement of $np$
spin-dependent total cross-section difference for beam
and target spin orientation transverse to the beam direction
$\Delta\sigma_T$
below 1200 MeV is plotted together with the phase-shift
solution SP95 from SAID~\cite{{arn1},{arn2}}.

Besides their general importance for the phase-shift analysis
(PSA) $\Delta\sigma_T$ values at the energies below 100 MeV
have showed their high sensitivity to the $^3S_1-^3D_1$
mixing parameter $\epsilon_1$ which is considered to be
ill-determined by several authors~\cite{bin}. This mixing
parameter measures the strength of a tensor component of the
$NN$ forces and can be determined from only a few other observables
which have a high
contribution of the coupled triplet. Taking into account the formulae
for the observables deduced in~\cite{bys} together with
experimental possibilities one comes to
the spin correlation coefficient $A_{yy}(\vartheta)$ (or $A_{00nn}$
in the Saclay four-subscript notation~\cite{bys})
 at 90$^{\rm o}$ c.m. angle (at this angle some amplitudes cancel
due to their antisymmetric behaviour) and the spin transfer parameter
$K_{y}^{y\prime}(\vartheta)$ (or $K_{0nn0}$). Analysis
of their measurements shows apparent discrepancies from the
potential model predictions (see subsect.~\ref{mixeps}).

        In this paper results from a new measurement of
$\Delta\sigma_T$   in neutron-proton scattering at 16.2 MeV
are presented.

\section{Experimental setup}

The measurement of $\Delta\sigma_T$ has been performed using the
classical transmission method, i.e. the relative difference in
attenuation of a
polarized neutron beam passing a polarized proton target
was measured.

\subsection{Polarized proton target} \label{ppt}

        In the present experiment the frozen spin polarized
proton target
has been used. The target is a complex device consisting of several blocks:
\begin {itemize}
\item High power $^3$He/$^4$He dilution refrigerator
\item Movable magnetic system
providing
a "warm" field and consisting of a superconducting solenoid and a
superconducting dipole holding magnet with a large aperture
\item Electronic
equipment for
providing proton polarization and measurement of its value.
\end{itemize}

1,2-propanediol
with a paramagnetic Cr(V) impurity was used as a target material.
Beads of this material approximately 2 mm diameter were
cooled to liquid nitrogen
temperature and placed inside a perforated teflon ampule 2 cm in
diameter and 6 cm
length. The ampule was then loaded into a horizontal mixing chamber of the
dilution
refrigerator using a lock device. A horizontal part of the refrigerator
containing almost 20 cm$^3$ of propanediol with a total mass of about
15 g was placed
in a neutron beamline.

     The hydrogen nuclei
 in the propanediol were polarized by a dynamic nuclear
orientation technique at a temperature 0.3-0.4 K in a strong highly
uniform magnetic
field (2.7 T) using 75 GHz hyperfrequency.
This magnetic field within the target volume
was produced by the superconducting solenoid
in the dynamic polarization regime.
Maximum values of the proton polarization obtained were 93\% and
98\% for
positive and negative polarizations, respectively. The time needed to achieve
80\% of maximum polarization is about one hour. The target polarization value
was calculated by comparison of the amplified
polarized proton NMR signal with a
thermal equilibrium NMR signal measured at about 1 K temperature in the
same magnetic field. The accuracy of the polarization determination was
approximately $\pm$3\%. This uncertainty is due to the evaluation
of integrated thermal equilibrium signal and to the measurement
of temperature.

        After achieving a high proton polarization the solenoid field
is decreased while the holding magnet field is increased. Finally, the
solenoid is removed from the vicinity of the target leaving it in 0.37 T
vertical magnetic field produced by the holding magnet. The latter
provides $\pm 50^{\rm{o}}$ aperture in the vertical plane and nearly
$360^{\rm{o}}$
aperture in the horizontal plane. The final temperature of the target
in the  frozen mode is around 20 mK. The spin relaxation
time for protons measured in these conditions was approximately
1000 hours for positive polarization and 300 hours
for negative polarization. As a result, a polarization degradation during
one continuous experimental run
of 10-12 hours was always insignificant.

        A more detailed description of the target complex is given
in~\cite{bor}.

\subsection{Polarized neutron beam} \label{pnb}

        The polarized neutron beam is produced as a secondary beam
via the $^{3}H(d,n)^{4}He$ reaction. An unpolarized deuteron beam with
$E_d=1.825$ MeV
from the Van de Graaff electrostatic accelerator  HV 2500 AN
strikes  a Ti-T target (2 mg/cm$^2$)
on a molybdenum backing at an angle of 45$^{\rm o}$.
To achieve a monoenergetic collimated neutron beam,
the associated particle method is used. The principle of the
method is as follows: knowing the incident projectile energy in
two-body reaction (using a thin target), the
energy and angle of emitted particle and recoil nucleus are
kinematically conjugated. In our case a collimated beam of
recoil alpha particles registered in the
charged particle detector at a definite angle is associated
with neutrons of known energy and
average angle emitted
in a narrow cone.
    The experimental setup is shown in Fig. 2.
For incident deuteron energies comparable to the recoil
alpha-particle energy, the charged particle detector suffers from
elastically scattered deuterons whose intensity relative to the number of
alpha-particles is  higher by several orders of magnitude
(due to the high Coulomb
cross-section).
To avoid this background we deflected parasitic deuterons from
the alpha-particles
using a magnetic separator.
Hence the recoil alpha-particles emitted at the laboratory
angle of 90$^{\rm o}$
to the primary deuteron beam together with the
deuterons elastically scattered at the same angle passed
through the perpendicular magnetic field of 0.5 T.
The silicon surface-barrier (SSB) detector ($8\times5$ mm$^2$)
was adjusted to the position corresponding to alpha-particles
curvature in the separator magnetic field, so it detected
only a small part of the original deuteron flux.
Neutrons associated with the alpha-particles, which registered in the
SSB detector, were emitted to the narrow cone (FWHM = 18 mm at 1 m
distance from the tritium target), which corresponds to the
aperture diameter in the shielding block ($0.6\times1.4\times1.4$
m$^{3}$).

In the experiment described here the associated neutron beam
with an energy $E_{n} = (16.2\pm 0.1)$  MeV
was emitted at an angle $\vartheta_{lab} = (62.0\pm 0.7)^{\rm o}$.
In spite of using the unpolarized incident deuteron beam the
outgoing neutron beam is partially polarized. Its polarization
in this energy region has been measured several
times~\cite{{lev},{per}}, but the
recent measurement in T\"ubingen \cite{koc} covers a large
range of energies and
angles with minimal error. Interpolating these results we have
obtained $P_b=(-13.5\pm1.4)$\% (see \ref{err})

The neutron production and detection system has been described in
a detail in~\cite{wilhelm}.

\subsection{Neutron detection system} \label{nds}

        The neutron beam incident on the polarized proton target is
monitored by two plastic scintillator detectors viewed via a
60 cm long light guide (to eliminate the magnetic field effects)
by
fast photomultipliers (PM) XP 2020 (monitor MON1:
$3\times20\times20$ mm$^{3}$, monitor MON2: $10\times30\times30$
mm$^{3}$).

Two liquid scintillator (NE-213) detectors DET1 and DET2
of cylindrical shape
($\o~40$ mm$\times60$ mm) were placed behind the polarized proton target.
These detectors were also mounted to
XP 2020 PM's.
All the detectors were located in the neutron beam axis.
A schematic diagram of the electronic circuit is shown in Fig. 3.
    The preamplifier attached to the semiconductor detector
gives two outputs:
fast timing signals and slow amplitude (proportional to energy) signals.

    The slow (energy) signal is amplified, fed into a linear gate and
stretcher, and then passed to the analog-to-digital converter ADC1. The fast
signal is
shaped in a constant-fraction discriminator (CFD) and fed along with the
signals from
the neutron detectors, also shaped in CFD's, to the coincidence control
unit. This
specially constructed programmable coincidence control unit with resolving time
equal to
100 ns allows registration of any combination of 8 input signals
defined by the control software while the registration of
any undefined combination is disabled. For our experiment the
coincidences of signals from one alpha and any (but only one)  neutron
detector (out of four) were enabled, any
coincidences between neutron detectors (coming from the multiple
neutron
scattering from one detector to another or from the background)
were disabled. The output digital signal from this unit gives information
about the type of coincidence, i.e. in our case the serial number of the
neutron detector
in coincidence with the alpha-particle detector.

Another output signal from the alpha-particle CFD is fed to the
time-to-amplitude converter (TAC) as a START
signal. STOP signal is derived from the fast summator fed with
CFD's signals corresponding to the neutron
detectors (this solution enables us to use only one TAC).
In order to minimize the dead time
of the TAC it is gated by a fast coincidence signal (formed from the same
signals as TAC) with resolving time equal to
full scale of TAC (100 ns) with a short delay time of response
(maximally 110 ns after the START signal). The output signal from TAC is
fed to ADC2. Both
the ADC's are gated by the output signal of the coincidence control.
Information from these two ADC's along with the coincidence type
number from the coincidence control unit is fed to
the computer via the CAMAC crate controller and parallel computer
interface.

\section{Experimental procedure}
\subsection{Formalism}

The expression for the nucleon-nucleon total cross-section for
polarized beam and target (deduced in ~\cite{bil} and \cite{phil}
 and discussed in ~\cite{bys}) can be written as

\begin{equation}
    \sigma_{tot} =\sigma_{0,tot}+\sigma_{1,tot}(\vec{P}_b
     \vec{ P_t})+\sigma_{2,tot}(\vec{P}_b \vec{k})(\vec{P}_t\vec{k})
\end{equation}
where $\vec P_b$ and $\vec{P}_t$ are the polarization vectors of the
beam and the target, respectively, and $\vec{k}$ is a unit vector in
the beam direction. For transverse and longitudinal spin
directions and complete polarizations
\begin{equation}
\mid\vec{P}_b\mid = \mid\vec{P}_t\mid=1
\end{equation}

we define the observables $\Delta \sigma_T$ and
$\Delta \sigma_L$ as

\begin{equation}
    \Delta\sigma_T=\sigma(\uparrow \downarrow)-\sigma(\downarrow
\downarrow)=-2\sigma_{1,tot}
\end{equation}

\begin{equation}
        \Delta\sigma_L=\sigma(\rightleftarrows)
                -\sigma(\rightrightarrows)
                =-2(\sigma_{1,tot}+\sigma_{2,tot})
%
%        \Delta\sigma_L=\sigma{\rightarrow \choose \leftarrow}
%                -\sigma{\rightarrow \choose \rightarrow}
%                =-2(\sigma_{1,tot}+\sigma_{2,tot})
\end{equation}

The relative difference in attenuation  of a
polarized neutron beam after passing
through the polarized proton target for parallel
and antiparallel spin orientations is

\begin{equation}
    \xi(c)=\frac{N_{d}(c)}{N_{mon}(c)}
\end{equation}
where $N_{d}$ and $N_{mon}$ are net areas under the time-of-flight
peak in the measured spectra from detector and monitor, respectively,
and $c$ denotes the spin orientation combination
($c=\uparrow \downarrow$ or $\downarrow \downarrow$).

Assuming  that the detector efficiencies $\eta_d$
and $\eta_{mon}$ are constant for all measurements (see subsect. 4.2)
and the degrees of beam and target polarizations are
$P_b$ and $P_t$, respectively,
we have
\begin{equation}
  N_d(\uparrow \downarrow)= \eta_d I_0 (\uparrow \downarrow) \exp[-\omega
      \sigma(\uparrow \downarrow)] =
   \eta_d I_0 (\uparrow \downarrow) \exp[-\omega
      (\sigma_{0}+\frac{1}{2}\Delta\sigma_T P_b P_t)]
\end{equation}
\begin{equation}
  N_d(\downarrow \downarrow)= \eta_d I_0 (\downarrow \downarrow) \exp[-\omega
      \sigma(\downarrow \downarrow)] =
   \eta_d I_0 (\downarrow \downarrow) \exp[-\omega
      (\sigma_{0}-\frac{1}{2}\Delta\sigma_T P_b P_t)]
\end{equation}
\begin{equation}
     N_{mon}(c)=
           \eta_{mon} I_0 (c)
\end{equation}
where $\omega$ is the number of protons per unit area of the target
and $I_0(c)$ is the integrated beam intensity.

Hence for $\Delta\sigma_T$  we finally have

\begin{equation}            \label{dst}
        \Delta\sigma_T=\frac{\ln(\xi(\downarrow \downarrow))-
               \ln(\xi(\uparrow \downarrow)) }
        {\omega P_{b}P_{t}}
\end{equation}

\subsection{Data collection}

The measurement was divided into 14 runs. During each run the target
polarization remained unchanged. Before the run the proton target
polarization was built up and measured.
Data consisting of ADC1 channel, ADC2 channel and the serial number of the
neutron detector were buffered and recorded on tape in
100-event-blocks together with other important information (total
counts in energy and time channel from the alpha detector as well
as from the individual neutron detectors, total time elapsed, etc.) During
the data acquisition some results were available from the
on-line monitoring program.

Immediately after each run the target polarization was measured and either
reversed or restored in its original magnitude for the next run.

The total data taking time was 91 hours for antiparallel and 85 hours
for parallel orientation of spins. During this time $8\times10^6$
antiparallel and $7\times10^6$ parallel net
$n-\alpha$ coincidences were recorded to tape. The typical
gross coincidence count rate was 50 s$^{-1}$ for approximately $10^4
$s$^{-1}$ count rates in both alpha and neutron channels. The
deuteron beam current on Ti-T target was kept below 5 $\mu$A.

\section{Results and discussion}

\subsection{$\Delta\sigma_T$ determination} \label{dstdet}

In the course of data reduction 2-parameter histograms (ADC1 vs.
ADC2) for each
neutron detector were created from tapes (see Fig. 4). A window in the
energy spectrum was set to eliminate elastically scattered
deuterons (see 2.3). Fig. 5 shows the typical charged particle
energy spectrum and Fig. 6 shows the %effect of the energy cut
%to the
neutron time-of-flight spectrum. Time resolution achieved
in this experimental setup with a residual magnetic field from
the polarized proton target was about 3 ns, the resolution with
the magnets off can be reduced down to 1.5 ns.
Applying this cut also reduced background of
random coincidences in the time-of-flight spectrum by a factor of
roughly 2. The remaining accidental background was linear in
a wide range of channels on both sides of the peak, so that a linear
approximation and subtraction could be used to calculate the net peak
area. The resulting area appeared to be fairly independent of the
variation of left and right peak borders.
Then four ratios of net areas from two detectors and two monitors
were calculated for each $10^5$ events. The stability control was
enabled by two additional ratios MON2:MON1 and DET2:DET1.

For both target spin orientations the weighted centroids of all
runs were calculated
\begin{equation}
        \langle \xi \rangle = \frac{\sum \frac{\xi_i
                P_i}{\sigma_i^2}}{\sum \frac{P_i}{\sigma_i^2}}
\end{equation}
\begin{equation}
        \langle P_t \rangle = \frac{\sum \frac{P_i}{\sigma_i^2}}
                {\sum \frac{1}{\sigma_i^2}}
\end{equation}
These centroid values of the target polarization $P_t$ and the
resulting ratio $\xi$ were then used to calculate the
spin-dependent total cross-section
difference
$\Delta\sigma_T$ from Eq.~\ref{dst}. The final
values were obtained as a weighted mean of the four relevant ratios.

        From our measurement we have obtained the result
\begin{equation}
\nonumber       \Delta\sigma_T=(-126\pm21\pm14)\quad \rm{mb}
\end{equation}
where the first uncertainty is the statistical error and
 the second uncertainty is due to systematic errors
(see subsect.~\ref{err}).
Comparison of our measured value with theoretical predictions as
well as with TUNL measurements is plotted in Fig. 7.

\subsection{Systematic errors and instrumental asymmetries}
\label{err}

To estimate the final uncertainty of $\Delta\sigma_T$ originating
from systematic effects several sources of errors were analysed.
They can be divided into two groups: polarization-dependent
effects which introduce a false asymmetry, and
polarization-independent effects included in the systematic error.

When evaluating polarization-dependent effects we did not
restrict ourselves only to the effects directly connected to the spins,
but we studied also side effects of polarization:
magnetic fields orientation, beam position, etc.

The use of the polarized proton target with identical
holding magnetic fields for both spin orientations is a great advantage
for this kind of measurement. To ensure this we measured the
magnetic
field intensity during the data collection near the
photomultiplier tubes. The monitoring showed that the relative changes
from one polarization to another as well as those during the run were
less than $3\cdot10^{-3}$, and as negligibly small were not taken
into account.

The displacement
of neutron detectors can be another possible source of false
asymmetries. Due to a non-zero analyzing power $A_y$
as well as the spin correlation coefficients $A_{yy}$ and
$A_{xx}$,
the left-right and up-down asymmetry for small angle $np$ elastic
scattering cross-section
exists.
When the detectors are placed symmetrically with respect
to the beam line the differences
will be averaged out, but any displacement will cause a non-zero
contribution to the measured spin-dependent cross-section.
Our calculations show that the maximal relative contribution does not
exceed  $4 \cdot 10^{-3}\rm{deg}^{-1} \approx 10^{-4}
\rm{mm}^{-1}$.
The tolerance of the detector adjustment is below 2 mm, so this source
of false asymmetry can be neglected.

Systematic error consists of several polarization-independent effects:
errors in determining the beam and target polarizations, error in target
density, polarization-independent effects of beam and detector geometry
and variation of detector efficiencies (due to instabilities of high
voltage, gains, thresholds, etc.).

The uncertainty of the neutron beam polarization manifests
itself as a major source of the
final error. We have taken the experimental values of polarization obtained
in 1991
in T\"ubingen~\cite{koc}, because they cover both the energy
and the angular region of our interest. The measured values
were fitted (using 2nd order polynomial for energy dependence
and Legendre polynomials for the angular dependence) and the resulting
value for $E_d=1.825$ MeV and $\vartheta=62^{\rm{o}}$ was taken.
Because $\chi^2/N_{d.o.f.}=1.2$ and our angle is close to the measured
angles $50^{\rm{o}}$ and $70^{\rm{o}}$, we kept the original absolute error
1.4 \%
which represents 10 \% relative error of $P_b$ and has a scale character.

The target polarization was measured with 3\% relative error (see subsect.
\ref{ppt}). The density of the target was determined by precision weighing
with 3\% relative error.

Geometrical displacement of detectors, beside its contribution to the
false asymmetry represents also a source of systematic error due to the
beam divergence and finite solid angle of the detectors.
Since detectors were not removed during the whole measuring period,
 this contribution was equal for both target spin orientations and this error
is a scale error. Calculations based on the beam profile
show relative contribution of 0.8\% for 2 mm displacement (equal to
the tolerance of detector adjustment).
The uncertainty originating from the beam position variations
($\approx 1$ mm) is about 0.1\%.

Use of scintillators attached to the photomultipliers
represents a considerable source of uncertainties (PM's are very
sensitive to the instability of high voltage applied, the
thresholds and gains of electronic modules used can float, etc.).
It should be noted that each PM was fed from an independent high voltage
supply.
We have performed a set of tests including the long-time monitoring
of high voltage supplies stability (within $\pm$1 V at 2000 V),
 runs with a ``dummy'' target without magnetic fields as well as
the analysis for two additional ratios MON2:MON1 and DET2:DET1 (see
subsect. \ref{dstdet}). All these tests yielded values consistent with
zero within error bars of 15 mb.

\subsection{The mixing parameter $\epsilon_1$} \label{mixeps}

Before we start the discussion on the influence of our measured value to the
determination of the mixing parameter $\epsilon_1$, we
describe briefly the experimental situation in the measurement
of $\epsilon_1$-related observables.
The $A_{yy}(\vartheta)$ measurement at
90$^{\rm o}$ c.m.
is presented by Sch\"oberl et al.~\cite{sch} for 13.7 MeV neutron
energy (Erlangen) and by Doll et al.~\cite{dol} for 19, 21 and 25 MeV
(Karlsruhe). The
spin transfer parameter $K_{y}^{y\prime}(133^{\rm o}_{c.m.})$ was measured at
25.8 and 17.4 MeV by Ockenfels et al.~\cite{{ock1},{ock2}} (Bonn),
while $\Delta\sigma_T$ was only recently measured  in TUNL in the 3.65-11.60
 MeV energy range~\cite{{wil},{tor2}}.
In all these works  the authors
performed at least basic phase shift analyses to determine the
value of $\epsilon_{1}$, in order to compare the experimental
results from different experiments. From the analyses (see Table
1 reprinted from \cite{wil})
it is evident that relative agreement exists between
individual experiments and potential model predictions, but there
are some indications of discrepancies between
13 and 20 MeV towards weaker tensor force. These indications were also
supported by
results published from TUNL~\cite{wil} for $E_n$=7.43, 9.57 and 11.6
MeV. In the meantime, the authors announced corrected
values (by about 20\%)~\cite{tor2} where the discrepancies are
not so apparent. However there are experimental results
supporting the hypothesis of lower tensor force around 15 MeV. This
is in contradiction to the result obtained by
the Basel group with an experiment performed at Villigen
\cite{{haff},{sick}},
where $\Delta\sigma_L$
has been measured at 66 MeV incident neutron energy and the
analysis made by Henneck \cite{hen}. In these works the authors
conclude that the tensor component of $NN$ potential below 100
MeV must be
stronger than predicted by the models.

As seen from Fig. 7 our new value of
$\Delta\sigma_T$ is in general agreement with potential model
predictions as well as with the TUNL measurements.
The comparison of other direct experimental results in the
field (see Introduction) is impossible so a phase-shift
analysis must be performed and resulting $\epsilon_1$ mixing parameter
values compared. Since it is not easy to perform a complete
PSA, most of the authors restrict themselves to varying only few
(or even one) phase parameter, while fixing the others at the
values from certain potential model or PSA, with a risk of
introducing ambiguities in the comparisons.
In the most extensive analysis of
$\epsilon_1$-oriented experiments below 30 MeV presented by
Wilburn~\cite{wil}, only $\epsilon_1$ was varied, while the remaining
phase parameters were taken from the Bonn B potential. We evaluated the
sensitivity of $\Delta\sigma_T$ to different phase shifts and
mixing parameters (using the full Bonn potential set), and the resulting
contributions
(see Table 2) justify the single parameter analysis to be performed
here.

The spin-dependent total cross-section difference
$\Delta\sigma_T(np)$ can be written as
\begin{equation}
           \Delta\sigma_T(np)   = \frac{1}{2}(
\Delta\sigma_T(I=0)  +\Delta\sigma_T(I=1))
\end{equation}
where I is the isospin,
or in terms of the phase-shifts in the Stapp
convention~\cite{{wildis},{tor}}

\begin{equation}
           \Delta\sigma_T   =
    \frac{\pi}{k^2} \left \{
  \begin{array}{l}
  [3\cos2\delta_{^1P_1} -\cos2\delta_{^3S_1} - 2\cos2\delta_{^3D_1}
   \\   +2\sqrt{2}\sin(\delta_{^3D_1}+\delta_{^3S_1})\sin2\epsilon_1
    +\ldots ]
    \\   +[ \cos2\delta_{^1S_0} - \cos 2\delta_{^3P_0}
    \\ +5\cos2\delta_{^1D_2}-2\cos2\delta_{^3P_2}-3\cos2\delta_{^3F_2}
   \\ +2\sqrt{6}\sin(\delta_{^3P_2}+\delta_{^3F_2})\sin2\epsilon_2 +
  \ldots ]
 \end{array} \right\}
\end{equation}

where $\delta_i$ is the phase-shift of a state i (in a spectroscopic
notation), $\epsilon_J$ is the mixing parameter of states with
total angular momentum $J$, and
$k$ is a neutron impulse in the centre-of-mass system. One can
see from these expressions that when the experimental difficulties
are overcome and the spin-dependent cross-section for $pp$ elastic
scattering is measured at this energy,
$\Delta\sigma_T(I=1)$ will be known and the number of phase
parameters to be varied will be reduced considerably.

For our value of $\Delta\sigma_T$ the resulting mixing parameter
is
\begin{equation}
\nonumber       \epsilon_1=(1.5\pm1.3)^{\rm{o}}
\end{equation}

The present situation in this energy region is displayed in Fig.
8,
where the results from Table 1 are plotted together with our
new result and with model predictions.

\section{Conclusion}

A measurement of the spin-dependent total cross-section
difference $\Delta\sigma_T$
for the scattering of polarized neutrons from  polarized protons
at 16 MeV has been made with the resulting value
\begin{equation}
\nonumber       \Delta\sigma_T=(-126\pm21\pm14)\quad \rm{mb}
\end{equation}

All effects possibly influencing the accuracy of the result have
been critically evaluated and the quoted uncertainty safely
encompasses all these effects.

A phase-shift analysis has been performed, varying the mixing
parameter $\epsilon_1$ while the other phase parameters
were kept constant and equal to the full Bonn potential model
predictions. This PSA gave the result
\begin{equation}
\nonumber       \epsilon_1=(1.5\pm1.3)^{\rm{o}}
\end{equation}

{}From this result
one can conclude that the presented values
do not support the hypothesis of local minimum of $\epsilon_1$
around 15 MeV, representing a much weaker tensor force than that
predicted (as indicated in \cite{{sch},{ock2},{wil}}).

Since the degree of our neutron beam polarization is rather low,
the error of its determination dominates the final error.
Any reduction of the uncertainty of $\Delta\sigma_T$ using our
current setup seems unfeasible.

One promising way
to obtain more precise experimental data for the phase-shift
analysis using the setup described here is to combine
more observables. Tornow et al.
\cite{tor3} have shown that the combination of the spin-dependent total
cross-section differences for transverse and longitudinal
polarizations $\Delta\sigma_T$ and $\Delta\sigma_L$ can reduce
the resulting inaccuracy in $\epsilon_1$ determinations. As far as
we are informed the $\Delta\sigma_L$ measurement is to be
performed in TUNL \cite{tor4}. Our experimental apparatus will be
modified to allow both beam and target spins to be
oriented in the longitudinal
direction. We are preparing the $\Delta\sigma_L$ measurement,
as we believe that new measurements are needed to
clarify the tensor contribution to the $NN$ potential in the low
energy region.

\section{Acknowledgment}

The authors express their gratitude to
J. Form\'anek, N.A. Russakovich and I. \'Ulehla for
their valuable contribution to the physics
program of this experiment and support and to F. Lehar for
careful reading of the manuscript and helpful comments.
Furthermore we thank M. Navr\'atilov\'a, V.G. Kolomiets and O.N.
Shchevelev for their assistance in carrying out the experiments.
        The authors gratefully acknowledge the financial
support from the Grant Agency of Czech Republic under the
registration No. 202/93/2426 for whom the work was performed.

\newpage
\pagestyle{empty}
%\begin{figure} \label{fexpset}
%\caption{
\begin{center}
{\bf Figure captions}
\end{center}
\begin{description}
\item[Fig.1] $\Delta\sigma_T$ measurements for the $np$ elastic
scattering below 1200 MeV with the phase-shift solution - vertical
axis shows $\frac{x}{\vert x \vert} \sqrt{\vert x \vert} $, where
$x=\Delta\sigma_T$ in mb (line -
SM95 solution from SAID~\cite{{arn1},{arn2}}, squares
- TUNL~\cite{{wil},{tor2}}, circle - this work, up triangles -
PSI~\cite{binz}, down triangles - Saturne II~\cite{lehar1})
\item[Fig.2] Top view of the experimental setup\
1. Deuteron beam 2. Ti-T target 3. Alpha-particle beam 4. Magnetic separator
5. Silicon surface-barrier detector 6. Neutron beam monitor MON1
7. Collimator and shielding 8. Neutron beam monitor MON2 9. Neutron beam
10. Polarized proton target 11. Neutron detector DET1
12. Neutron detector DET2
%\end{figure}[p]
%
%\begin{figure}
\item[Fig.3] Schematic diagram of the electronic circuit
(for further description see
subsect. 2.2 and 2.3)\
\begin{flushleft}
SSB: Silicon surface-barrier detector \\
PA: Preamplifier Canberra\\
SA: Spectroscopic amplifier Tennelec TC-244\\
S: Liquid scintillator NE-213\\
PMT: Photomultiplier tube Philips XP-2020\\
LG\&S: Linear gate \& stretcher Ortec 542\\
CFD: Constant fraction discriminator Tennelec TC-454\\
TAC: Time to amplitude converter Tennelec TC-816A\\
ADC: Analog to digital converter Tesla NL-2320\\
FC: Fast coincidence (Nuclear Centre)\\
CC: Coincidence control (Nuclear Centre)\\
SUM: Fast summator (Nuclear Centre)\\
MPU: Multiparameter unit (Nuclear Centre)\\
Variable delay Polon 1506\\
CAMAC crate controller Tesla NL-2106\\
\end{flushleft}
\item[Fig.4] Scatterplot of 2-parameter histogram: x-axis
corresponds to the neutron detector time-of-flight
 (ADC2) spectrum, y-axis corresponds to
the energy spectrum of the SSB detector (ADC1).
\item[Fig.5] Energy spectrum of the SSB detector: solid line --
 original spectrum, hatched area -- spectrum
gated with $\alpha-n$
coincidences (the deuterons are well suppressed, see subsect.~ \ref{pnb})
\item[Fig.6] Time-of-flight spectrum from the neutron detector
\item[Fig.7] Comparison of existing $\Delta\sigma_T$ measurements
and potential model predictions (circles - TUNL
\cite{{wil},{tor2}}, triangle - this work, lines: solid - Bonn,
dashed - Paris, dotted - Nijmegen)
\item[Fig.8] Values of $\epsilon_1$ analysed from available data
(full squares -  TUNL \cite{{wil},{tor2}}, diamond - Erlangen
\cite{sch}, open square - this work, open triangles - Karlsruhe
\cite{dol}, full triangles - Bonn \cite{{ock1},{ock2}}, lines:
solid - Bonn potential, dashed - Nijmegen, dotted - Paris,
dashed-dotted - Low energy (0-400 MeV) solution VZ40 from SAID
\cite{{arn1},{arn2}}
\end{description}
\newpage
\begin{table} \label{tab1}
\begin{center}
\caption{Values of $^3 D_1-^3 S_1$ mixing parameter $\epsilon_1$
from presently available data} %(from {\cite{wil})}
\vspace {.3cm}
   \begin{tabular}{ccccccr}
    \hline
     $E_n$ (MeV)&$\epsilon_1$ (degrees)&Observable & Ref.\\
      \hline
   5.1 & 0.41$\pm$0.22 & $\Delta\sigma_T$ & \cite{{wil},{tor2}}\\
        7.4 & 0.54$\pm$0.43 & $\Delta\sigma_T$ & \cite{{wil},{tor2}}\\
        9.6 & 1.32$\pm$0.51 & $\Delta\sigma_T$ & \cite{{wil},{tor2}}\\
        11.6 & 1.50$\pm$0.64 & $\Delta\sigma_T$ & \cite{{wil},{tor2}}\\
        13.7 & -0.16$\pm$0.50 & $A_{yy}$ & \cite{sch}\\
        17.4 & -0.94 $\pm$1.11 & $K_{y}^{y\prime}$ & \cite{ock2}\\
        19.0 &  1.20$\pm$0.94 & $A_{yy}$ & \cite{dol}\\
        22.0 &  1.46$\pm$0.66 & $A_{yy}$ & \cite{dol}\\
        25.0 &  2.64$\pm$0.68 & $A_{yy}$ & \cite{dol}\\
        25.8 &  2.60 $\pm$0.40 & $K_{y}^{y\prime}$ & \cite{ock1}\\
       \hline
    \end{tabular}
  \end{center}
\end{table}
\begin{table} \label{tab2}
\begin{center}
\caption{Sensitivity of $\Delta \sigma_T$ to changes in
phase-shifts and mixing parameters in relative units (for the
full Bonn set)}
\vspace{.3cm}
   \begin{tabular}{ccccccccccccr}
    \hline
     $\delta_{^1S_0}$&$\delta_{^3P_0}$&$\delta_{^1P_1}$ &
$\delta_{^3P_1}$ & $\delta_{^3S_1}$ & $\epsilon_1$&
     $\delta_{^3D_1}$&$\delta_{^1D_2}$&$\delta_{^3D_2}$ &
$\delta_{^3P_2}$ & $\delta_{^3F_2}$ & $\epsilon_2$ \\
      \hline
0.19&0.05&0.11&0.00&0.01&0.57&0.02&0.01&0.00&0.01&0.01&0.03 \\
        \hline
    \end{tabular}
  \end{center}
\end{table}

\pagebreak
\begin{figure}[p]
%\label{fig:jeden}
\begin{center}
  \mbox{\epsfig{file=fig01.ps,height=18cm, width=15cm,bbllx=0cm,
      bblly=0cm, bburx = 21cm, bbury = 25cm}}
\end{center}
\vspace*{-2cm}
\begin{center} Figure 1 \end{center}
%\caption{Figure 1}
\end{figure}

\pagebreak

\begin{figure}[p]
%\label{fig:tri}
\begin{center}
  \mbox{\epsfig{file=fig02.ps,height=18cm,width=15cm,bbllx=0cm,
      bblly=0cm, bburx = 21cm, bbury = 25cm}}
\end{center}
\vspace*{-2.4cm}
\begin{center} Figure 2 \end{center}
%Figure 5
\end{figure}
\pagebreak

\begin{figure}[p]
%\label{fig:tri}
\begin{center}
  \mbox{\epsfig{file=fig03.ps,height=18cm,width=15cm,bbllx=0cm,
      bblly=0cm, bburx = 21cm, bbury = 25cm}}
\end{center}
\vspace*{-2.4cm}
\begin{center} Figure 3 \end{center}
%Figure 5
\end{figure}
\pagebreak

\begin{figure}[p]
%\label{fig:tri}
\begin{center}
  \mbox{\epsfig{file=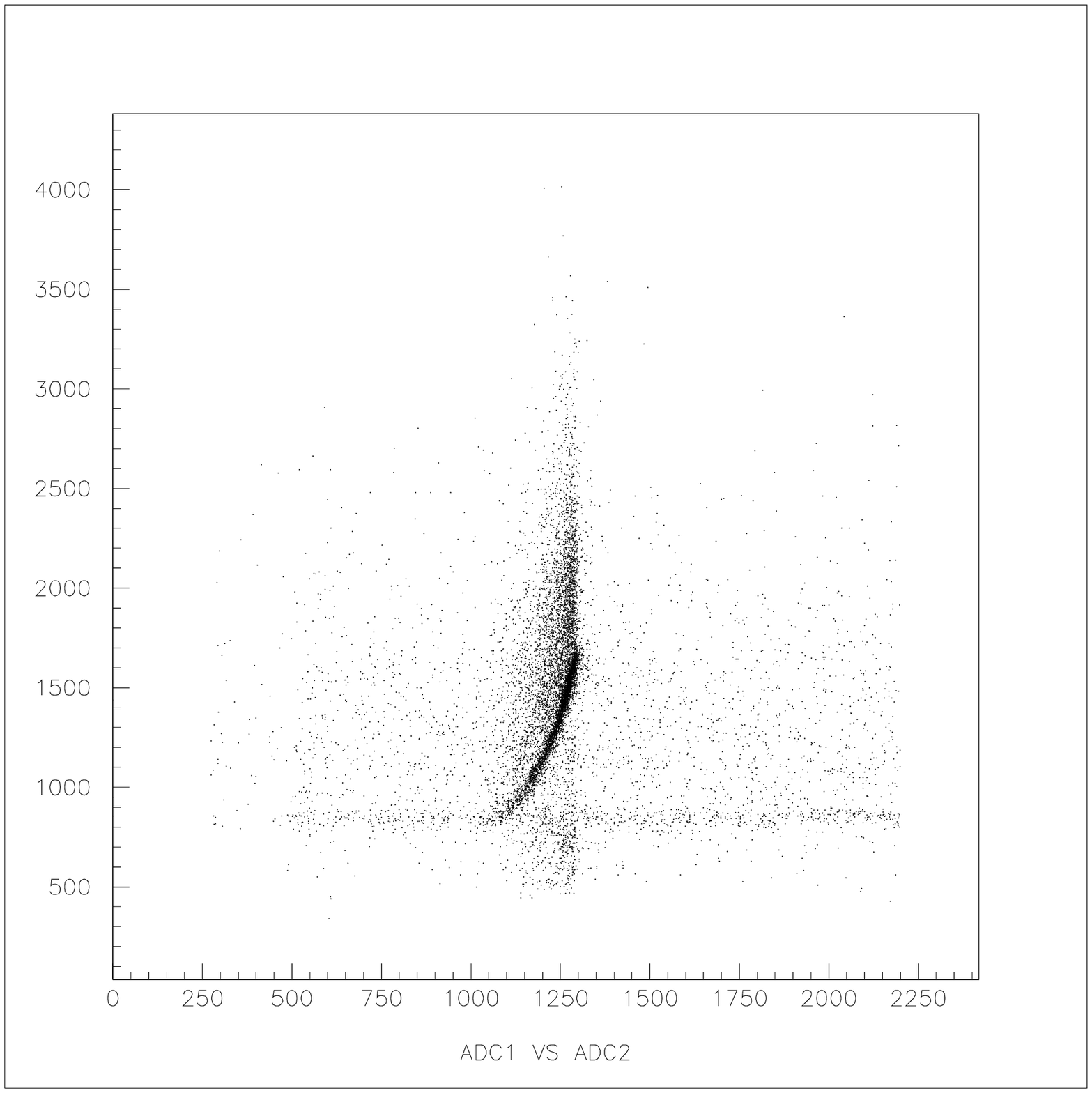,height=18cm,width=15cm,bbllx=0cm,
      bblly=0cm, bburx = 21cm, bbury = 25cm}}
\end{center}
\vspace*{-2.4cm}
\begin{center} Figure 4 \end{center}
%Figure 5
\end{figure}
\pagebreak

\begin{figure}[p]
%\label{fig:tri}
\begin{center}
  \mbox{\epsfig{file=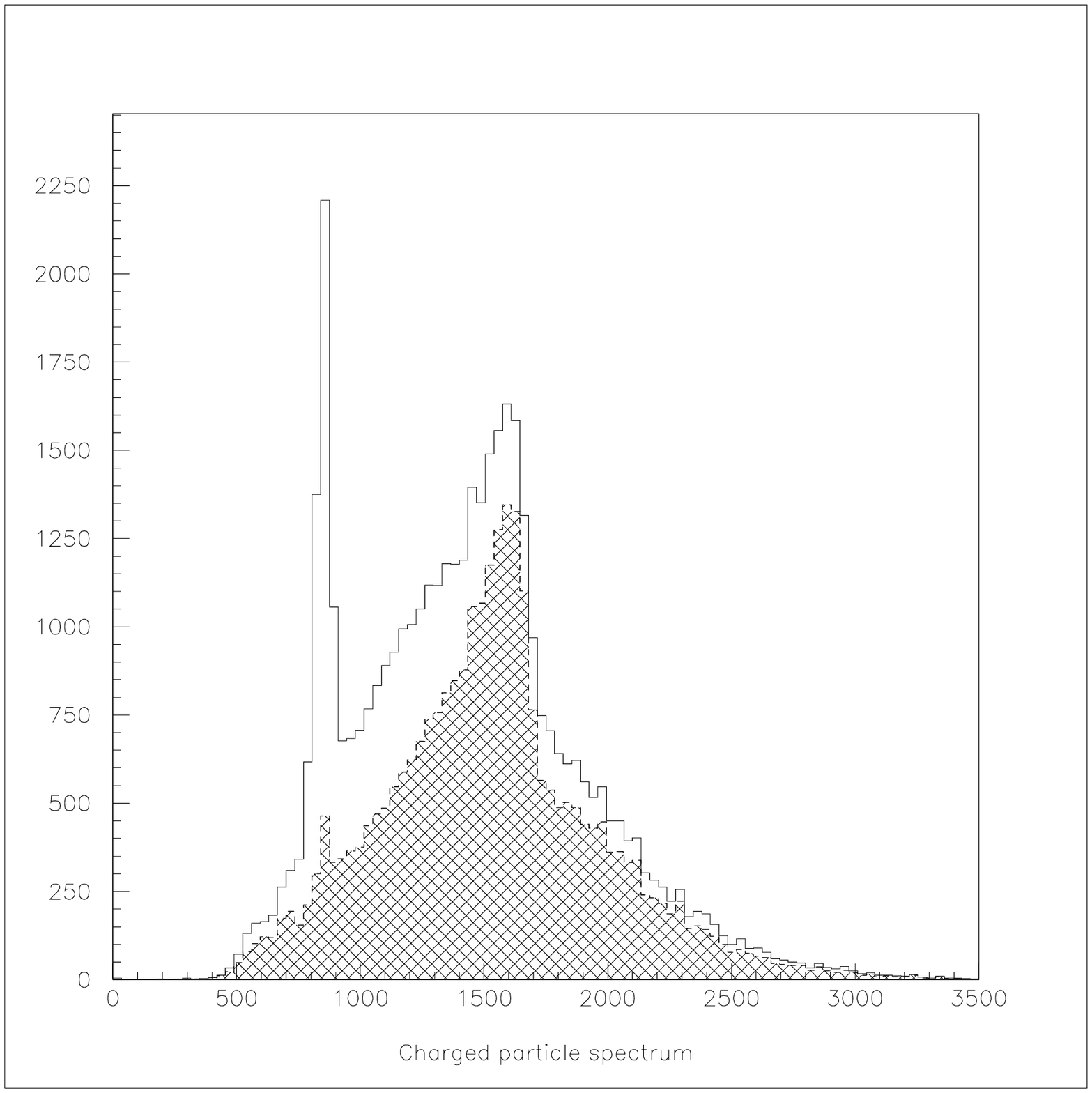,height=18cm,width=15cm,bbllx=0cm,
      bblly=0cm, bburx = 21cm, bbury = 25cm}}
\end{center}
\vspace*{-2.4cm}
\begin{center} Figure 5 \end{center}
%Figure 5
\end{figure}
\pagebreak

\begin{figure}[p]
%\label{fig:tri}
\begin{center}
  \mbox{\epsfig{file=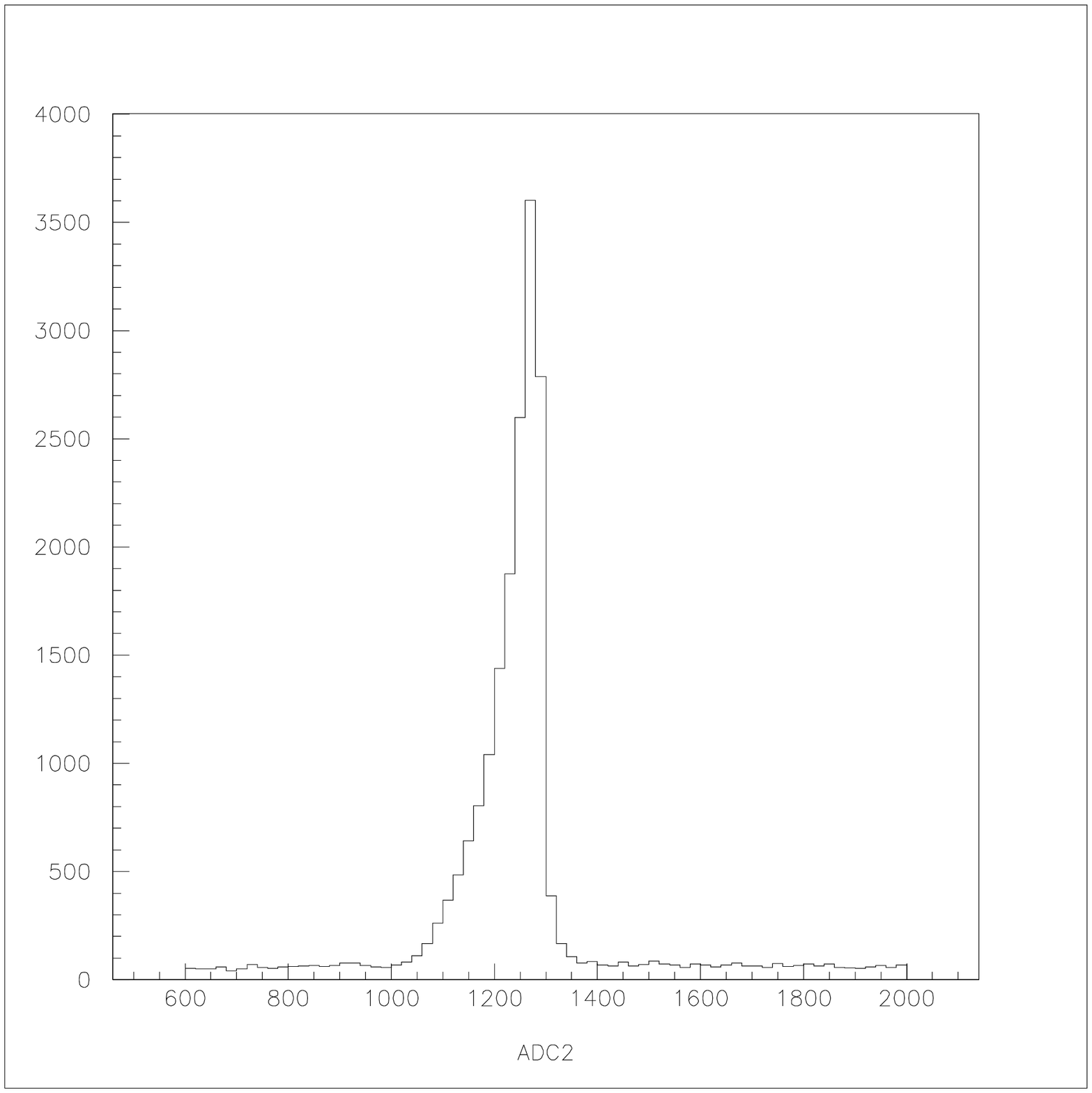,height=18cm,width=15cm,bbllx=0cm,
      bblly=0cm, bburx = 21cm, bbury = 25cm}}
\end{center}
\vspace*{-2.4cm}
\begin{center} Figure 6 \end{center}
\end{figure}
\pagebreak

\begin{figure}[p]
%\label{fig:tri}
\begin{center}
  \mbox{\epsfig{file=fig07.ps,height=18cm,width=15cm,bbllx=0cm,
      bblly=0cm, bburx = 21cm, bbury = 25cm}}
\end{center}
\vspace*{-2.4cm}
\begin{center} Figure 7 \end{center}
%Figure 5
\end{figure}
\pagebreak

\begin{figure}[p]
%\label{fig:tri}
\begin{center}
  \mbox{\epsfig{file=fig08.ps,height=18cm,width=15cm,bbllx=0cm,
      bblly=0cm, bburx = 21cm, bbury = 25cm}}
\end{center}
\vspace*{-2.4cm}
\begin{center} Figure 8 \end{center}
%Figure 5
\end{figure}

\end{document}